\documentstyle[11pt]{article}
\begin{document}

\title{\textbf{Holographic $f(T)$-gravity model
with power-law entropy correction}}

\author{K. Karami$^{1}$\thanks{KKarami@uok.ac.ir} ,
S. Asadzadeh$^{1}$, A. Abdolmaleki$^{1}$, Z. Safari$^{2}$\\
$^{1}$\small{Department of Physics, University of Kurdistan,
Pasdaran St., Sanandaj, Iran}\\$^{2}$\small{Research Institute for
Astronomy and Astrophysics of Maragha (RIAAM), Maragha, Iran}}

\maketitle

\begin{abstract}
Using a correspondence between the $f(T)$-gravity with the power-law
entropy corrected version of the holographic dark energy model, we
reconstruct the holographic $f(T)$-gravity model with power-law
entropy correction. We fit the model parameters by using the latest
observational data including type Ia supernovea, baryon acoustic
oscillation, cosmic microwave background, and Hubble parameter data.
We also check the viability of our model using a cosmographic
analysis approach. Using the best-fit values of the model, we obtain
the evolutionary behaviors of the effective torsion equation of
state parameter of the power-law entropy corrected holographic
$f(T)$-gravity model as well as the deceleration parameter of the
universe. We also investigate different energy conditions in our
model. Furthermore, we examine the validity of the generalized
second law of gravitational thermodynamics. Finally, we point out
the growth rate of matter density perturbation in our model. We
conclude that in power-law entropy corrected holographic
$f(T)$-gravity model, the universe begins a matter dominated phase
and approaches a de Sitter regime at late times, as expected. It
also can justify the transition from the quintessence state to the
phantom regime in the near past as indicated by recent observations.
Moreover, this model is consistent with current data, passes the
cosmographic test and fits the data of the growth factor well as the
$\Lambda$CDM model.
\end{abstract}


\clearpage

\clearpage
\section{Introduction}\label{intro}

Astronomical data from the type Ia supernovae (SNeIa), cosmic
microwave background (CMB) and baryon acoustic oscillation (BAO),
etc., have implied that the current expansion of the universe is
accelerating \cite{Riess1}. The proposals that have been put forth
to explain this observed phenomenon can basically be classified into
two categories. One is to introduce some unknown matters with
negative pressure called ``dark energy'' (DE) in the framework of
Einstein's general relativity (for reviews on DE, see e.g.
\cite{Padmanabhan}). Another alternative to account for the current
accelerating cosmic expansion is to modify the gravitational theory
called ``dark gravity'' (see e.g. \cite{Tsujikawa} for a review on
modified gravity).

Among the many dynamical DE models, a class with feature of quantum
gravity looks very special and attractive. Such class of models,
usually called holographic DE (HDE) \cite{Li}, has been motivated
from the holographic principle \cite{Suss1}. The energy density of
the HDE is given by
\begin{equation}
\rho_{\Lambda}=3c^2M_P^2L^{-2},\label{rhoHDE}
\end{equation}
where $c$ is a numerical constant \cite{Li}. The HDE models have
been studied widely in the literature
\cite{Guberina,Bekenstein,Li55,Li6,Enqvist,Elizalde2,Guberina1,Guberina2,Karami1}.
The derivation of the HDE density depends on the entropy-area
relationship $S_{\rm BH} = A/(4G)$, with $A\sim L^2$ which is the
area of horizon. This definition for the entropy in the presence of
quantum effects can be modified. The quantum correction to the
horizon entropy reads \cite{Saurya}
\begin{equation}
S_{\rm A}=\frac{A}{4G}\left[1-K_{\alpha}
A^{1-\frac{\alpha}{2}}\right],\label{ec}
\end{equation}
where $\alpha$ is a dimensionless parameter and
\begin{equation}
K_\alpha=\frac{\alpha(4\pi)^{\frac{\alpha}{2}-1}}{(4-\alpha)r_c^{2-\alpha}},
\end{equation}
here $r_c$ is the crossover scale. The second term in Eq. (\ref{ec})
can be regarded as a power-law correction to the Bekenstein-Hawking
entropy-area relation $S_{\rm A}=S_{\rm BH}=A/(4G)$, resulting from
entanglement, when the wave-function of the field is chosen to be a
superposition of ground state and exited state
\cite{Saurya,Shankaranarayanan1,Shankaranarayanan2}.

Following the derivation of the HDE \cite{Guberina} and taking the
relation (\ref{ec}) into account, the HDE density will be modified.
The result yields the power-law entropy-corrected HDE (PLECHDE)
density as \cite{SJ}
\begin{equation}\label{rhoPLECHDE}
\rho _{\Lambda}=3c^2M_{P}^{2}L^{-2}-\beta M_{P}^{2}L^{-\alpha},
\end{equation}
where $\beta$ is a dimensional constant. Karami et al. \cite{KS}
investigated the validity of the generalized second law of
gravitational thermodynamics on the apparent horizon for the
power-law corrected entropy-area relation (\ref{ec}) and concluded
that the GSL is satisfied from the past to the present for
$\alpha<2$ and it is violated in the future. In section \ref{Obs},
we will use the current observational data to constrain the
parameter $\alpha$. Karami et al. \cite{KS}, interestingly enough,
also found that for the PLECHDE model (\ref{rhoPLECHDE}) which is
the power-law entropy-corrected version of the HDE model
(\ref{rhoHDE}), the identification of IR cut-off with Hubble
horizon, $L=H^{-1}$, can lead to a phantom accelerating universe.
This is in contrast to the ordinary HDE where its equation of state
parameter behaves like the dust (or dark) matter if one chooses
$L=H^{-1}$.

In the framework of modified gravity, recently a new dark gravity
theory, namely the so-called $f(T)$ theory, attracted much attention
in the community, where $T$ is the torsion scalar
\cite{Ferraro,bengochea}. It has been demonstrated that the $f(T)$
theory can not only explain the present cosmic acceleration with no
need of DE \cite{bengochea}, but also provide an alternative to
inflation without an inflaton \cite{Ferraro}. $f(T)$ theory is based
on the old idea of teleparallel gravity (TG) \cite{Einstein}, in
which the Weitzenb\"{o}ck connection rather than the Levi-Civita
connection is used. As a result, the space-time has only torsion and
thus is curvature-free. Although TG is closely related to standard
GR, differing only in terms involving total derivatives in the
action, i.e. boundary terms \cite{Moller}, there are some
fundamental conceptual differences between them. According to GR,
gravity curves the space-time and shapes the geometry. In TG however
torsion does not shape the geometry but instead acts as a force. It
means that in TG there are no geodesic equations but there are force
equations much like the Lorentz force in electrodynamics
\cite{Hayashi}. In the literature, several gravitational theories
with torsion were proposed (see e.g. \cite{CapReview} for a good
review on extended theories of gravity).

$f(T)$ theory is obtained by extending the action of TG in analogy
to the $f(R)$ theory. An important advantage of $f(T)$ theory is
that its field equations are second order as opposed to the fourth
order equations of $f(R)$ gravity \cite{WufT}. This feature has led
to a rapidly increasing interest in the literature. Numerous
features of theoretical and observational interests have been
studied in this gravity model already including some viable
phenomenological $f(T)$ models \cite{Linder}, observational
constraints \cite{Wu}, cosmological perturbations and growth factor
of matter perturbations \cite{Saridakis}, matter stability
\cite{Nozari}, Birkhoff's theorem \cite{Meng}, Static solutions with
spherical symmetry \cite{Wang}, cosmography
\cite{CapCosmo,CapCosmoRevised}, and thermodynamical description of
$f(T)$-gravity \cite{Miao,KA}. For reviews on other aspects of TG
and $f(T)$-gravity, see \cite{Myrzakulov,Yang,Sadjadi}.

In the present work, our aim is to reconstruct a $f(T)$-gravity
model without resorting to any additional DE, that is, considering
that the PLECHDE is effectively described by the modification of the
gravity with respect to the TG. To do so, in section \ref{fT}, we
briefly review the $f(T)$-gravity in a spatially flat FRW universe
filled only with the pressureless matter. In section \ref{PLECfT},
we reconstruct a $f(T)$ model according to the evolution of PLECHDE
density. In section \ref{Obs}, we fit this model and give the
constraints on model parameters, with current observational data
including SNeIa, CMB, BAO and observational Hubble data (OHD). In
section \ref{numeric}, we give the numerical results. In section
\ref{cosmo}, we check the viability of our model using the
cosmographic analysis method. In section \ref{GSLaw}, the validity
of the generalized second law of gravitational thermodynamics for
our $f(T)$ model is examined. In section \ref{str}, we study the
growth of structure formation in our model. Section \ref{conc} is
devoted to conclusions.

\section{The $f(T)$ theory of gravity}\label{fT}

The modified teleparallel action of a generic $f(T)$ model with the
matter Lagrangian $L_m$ is \cite{Ferraro,bengochea}
\begin{equation}
I =\frac{1}{2k^2}\int {\rm d}^4
x~e~\Big[f(T)+L_m\Big],\label{action}
\end{equation}
where $k^2=M_P^{-2}=8\pi G$, $e={\rm det}(e^i_{\mu})=\sqrt{-g}$ and
$T$ is the torsion scaler. Here $e^i_{\mu}$ is the vierbein field
which uses as dynamical object in TG.

The modified Friedmann equations in the case of $f(T)$-gravity for
the spatially flat FRW universe are given by \cite{WufT,KA}
\begin{equation}
\frac{3}{k^2}H^2=\rho_m+\rho_T,\label{fT11}
\end{equation}
\begin{equation}
\frac{1}{k^2}(2\dot{H}+3H^2)=-(p_m+p_T),\label{fT22}
\end{equation}
where
\begin{equation}
\rho_T=\frac{1}{2k^2}(2Tf_T-f-T),\label{roT}
\end{equation}
\begin{equation}
p_T=-\frac{1}{2k^2}[-8\dot{H}Tf_{TT}+(2T-4\dot{H})f_T-f+4\dot{H}-T],\label{pT}
\end{equation}
\begin{equation}
T=-6H^2,\label{T}
\end{equation}
and $H=\dot{a}/a$ denotes the Hubble parameter. Here $\rho_m$ and
$p_m$ are the energy density and pressure of the matter inside the
universe, respectively. Also $\rho_T$ and $p_T$ are the torsion
contributions to the energy density and pressure. The energy
conservation laws are given by
\begin{equation}
\dot{\rho}_m+3H(\rho_m+p_m)=0,
\end{equation}
\begin{equation}
\dot{\rho}_T+3H(\rho_T+p_T)=0.\label{ecT}
\end{equation}
By using Eqs. (\ref{roT}) and (\ref{pT}), one can define the
effective torsion equation of state (EoS) parameter as
\cite{WufT,KA}
\begin{equation}
\omega_T=\frac{p_T}{\rho_T}=-1+\frac{8\dot{H}Tf_{TT}+4\dot{H}f_T-4\dot{H}}{2Tf_T-f-T}.\label{omegaT}
\end{equation}
With the help of Eqs. (\ref{fT11}), (\ref{roT}) and (\ref{T}) one
can get
\begin{equation}
\rho_m=\frac{1}{16\pi G}(f-2Tf_T).\label{rhom}
\end{equation}
For the pressureless matter, i.e. $p_m=0$, from Eqs. (\ref{fT11}) to
(\ref{pT}) one can obtain
\begin{equation}
\dot{H}=-\frac{4\pi G\rho_m}{f_T+2Tf_{TT}}.\label{Hdot}
\end{equation}
Inserting Eq. (\ref{rhom}) into (\ref{Hdot}) and using
$\dot{T}=-12H\dot{H}$ gives
\begin{equation}
\dot{T}=3H\left(\frac{f-2Tf_{T}}{f_T+2Tf_{TT}}\right).\label{Tdot}
\end{equation}
Using the above relation, the effective EoS parameter (\ref{omegaT})
yields
\begin{equation}
\omega_T=-\frac{f/T-f_T+2Tf_{TT}}{(f_T+2Tf_{TT})(f/T-2f_T+1)}.\label{omegaT2}
\end{equation}
It is also interesting to study the behavior of the deceleration
parameter defined as
\begin{equation}
q=-1-\frac{\dot{H}}{H^2},\label{q1}
\end{equation}
which can be compared with the observations. Using Eqs. (\ref{T})
and (\ref{Tdot}) the deceleration parameter (\ref{q1}) leads to
\begin{equation}
q=2\left(\frac{f_T-Tf_{TT}-\frac{3f}{4T}}{f_T+2Tf_{TT}}\right).\label{q2}
\end{equation}

\section{Power-law entropy corrected holographic $f(T)$-gravity
model}\label{PLECfT}

The dark torsion contribution in $f(T)$-gravity can justify the
observed acceleration of the universe without resorting to the DE.
This motivates us to reconstruct a $f(T)$-gravity model according to
the PLECHDE model. Following \cite{KS,SJ}, the PLECHDE density with
the Hubble IR cut-off $L=H^{-1}$ is given by
\begin{equation}
\rho_{\Lambda}=\frac{3c^2}{k^2}H^2-\frac{\beta}{k^2}H^\alpha,\label{ro
H}
\end{equation}
where $c$, $\alpha$ and $\beta$ are constants. For $\beta=0$ the
above equation transforms to the well-known HDE density with the
Hubble horizon \cite{Li}.

Replacing $H=(-\frac{T}{6})^{1/2}$ into (\ref{ro H}) yields
\begin{equation}
\rho_\Lambda=-\frac{c^2}{2k^2}T-\frac{\gamma}{2k^2}(-T)^\frac{\alpha}{2},\label{ro
H R}
\end{equation}
where
\begin{equation}
\gamma=\frac{2\beta}{6^{\alpha/2}}.\label{alpha}
\end{equation}
Equating (\ref{roT}) with (\ref{ro H R}), i.e.
$\rho_T=\rho_{\Lambda}$, gives the following differential equation
\begin{equation}
2Tf_T-f-(1-c^2)T+\gamma(-T)^\frac{\alpha}{2}=0.\label{dif eq1}
\end{equation}
Solving Eq. (\ref{dif eq1}) yields the power-law entropy corrected
holographic (PLECH) $f(T)$-gravity model as
\begin{equation}
f(T)=\epsilon
\sqrt{-T}+(1-c^2)T+\frac{\gamma}{1-\alpha}(-T)^\frac{\alpha}{2},\label{fHDE}
\end{equation}
where $\epsilon$ is an integration constant that can be determined
from a boundary condition. Following \cite{CapCosmo} to recover the
present day value of Newtonian gravitational constant we need to
have
\begin{equation}
f_T(T_0)=1,\label{f_T0}
\end{equation}
where $T_0=-6H_0^2$ is the torsion scalar at the present time.
Applying the above boundary condition to the solution (\ref{fHDE})
one can obtain
\begin{equation}
\epsilon=-2\sqrt{6}H_{0} \left[c^2 + \frac{\gamma
\alpha}{2(1-\alpha)} (6H_{0}^{2})^{\frac{\alpha}{2}-1}
\right].\label{ep}
\end{equation}

Note that the parameter $\gamma$ can be obtained by inserting Eq.
(\ref{fHDE}) into the modified Friedmann equation (\ref{fT11}).
Solving the resulting equation for the present time gives
\begin{equation}
\gamma=\frac{\Omega_{m_{0}} + c^2 -
1}{(6H_{0}^{2})^{\frac{\alpha}{2}-1}},\label{delta}
\end{equation}
Replacing Eq. (\ref{delta}) into (\ref{ep}) yields
\begin{equation}
\epsilon=\sqrt{6}H_{0}\left[\frac{\alpha(\Omega_{m_0}-1)+c^2(2-\alpha)}{\alpha-1}\right].\label{ep2}
\end{equation}
Using Eqs. (\ref{fT11}), (\ref{roT}), (\ref{fHDE}) and (\ref{delta})
one can obtain the dimensionless Hubble parameter
$E(z;\textbf{p})=H(z;\textbf{p})/H_0$ as
\begin{equation}
E^{2}(z;\textbf{p}) = \Omega_{m_{0}}(1 + z)^3 + c^2
E^{2}(z;\textbf{p}) + (1 - \Omega_{m_{0}} -
c^2)E^{\alpha}(z;\textbf{p}).\label{eqE}
\end{equation}
where $\Omega_{m_0}h^2=0.1352 \pm 0.0036~(68\%~{\rm CL})$ is the
present value of the dimensionless matter energy density and
$H_{0}=70.2 \pm 1.4 {\rm~km~s^{-1}~Mpc^{-1}}$(68\% CL) is the
present Hubble constant which has been updated in the 7-year WMAP
(WMAP7) data \cite{Komatsu}. Also \textbf{p} indicate model
parameters. Thus, throughout this work we fix the dimensionless
matter energy density and Hubble parameters at
$\Omega_{m_0}h^2=0.1352$ and $H_0 = 70.2$. With $\Omega_{m_0}$ and
$H_0$ being determined by independent measurements, in the next
section we will use the cosmic observations to constrain the PLECH
$f(T)$-gravity model parameters $\textbf{p}=(\alpha,c^2)$.

Inserting Eq. (\ref{fHDE}) into (\ref{omegaT2}) gives the EoS
parameter of the torsion contribution as
\begin{equation}
\omega_{T}=\frac{\gamma (\alpha-2)
(-T)^{\frac{\alpha}{2}+1}}{\Big[c^2 T +
\gamma(-T)^{\frac{\alpha}{2}}\Big]\Big[2(c^2 - 1)T+ \alpha \gamma
(-T)^{\frac{\alpha}{2}}\Big]}.\label{wHDE}
\end{equation}
Inserting Eq. (\ref{fHDE}) into (\ref{q2}), the deceleration
parameter takes the form
\begin{equation}
q=\frac{(c^2 - 1)T- \gamma(\alpha-3)(-T)^{\frac{\alpha}{2}}}{2(c^2 -
1)T+ \alpha \gamma (-T)^{\frac{\alpha}{2}}}.\label{q3}
\end{equation}

\section{Observational constraints}\label{Obs}

Here, we fit the free parameters of the PLECH $f(T)$-gravity model
by using the recent observational data including SNeIa, BAO, CMB and
OHD.

\subsection{Type Ia Supernovae (SNeIa)}

SNeIa can be used to directly measure the expansion rate of the
universe up to high redshift. We use the Union2.1 compilation
\cite{union} containing 580 SNeIa. It is an updated version of the
Union2 compilation \cite{Amanullah}. Constraints from the SNeIa data
can be obtained by fitting the distance modulus $\mu(z)$. A distance
modulus can be calculated as \cite{Pietro,Nesseris}
\begin{equation}
\mu_{\rm th}(z)=5\log_{10}D_{\rm L}(z)+\mu_{0},
\end{equation}
where $\mu_{0}=42.38-5\log_{10}h$ and $h$ is the Hubble constant
$H_0$ in units of $100~\rm{km~s^{-1}~Mpc^{-1}}$. Also the
Hubble-free luminosity distance $D_{\rm L}(z)$ for the flat universe
is given by
\begin{equation}
 D_{\rm L}(z)=(1+z)\int_{0}^{z}\frac{dz'}{E(z';\textbf{p})}.\label{DL-SNeIa}
 \end{equation}
Using SNeIa data, theoretical model parameters can be determined by
minimizing \cite{Pietro,Nesseris}
\begin{equation}
\tilde{\chi}_{\rm SN}^{2}=A-\frac{B^{2}}{C},
\end{equation}
where
\begin{equation}
A=\sum_{\rm i=1}^{580}[\mu_{\rm obs}(z_{\rm i})-\mu_{\rm th}(z_{\rm
i})]^{2}/\sigma_{\rm i}^{2},
\end{equation}
\begin{equation}
B=\sum_{\rm i=1}^{580}[\mu_{\rm obs}(z_{\rm i})-\mu_{\rm th}(z_{\rm
i})]/\sigma_{\rm i}^{2},
\end{equation}
\begin{equation}
C=\sum_{\rm i=1}^{580}1/\sigma_{\rm i}^{2},
\end{equation}
and $\sigma_{\rm i}$ stands for the $1\sigma$ uncertainty associated
to the $i$th data point.

\subsection{Baryon Acoustic Oscillations (BAO)}

BAO can be traced to pressure waves at the recombination epoch
generated by cosmological perturbations in the primeval
baryon-photon plasma. They have been revealed by a distinct peak in
the large scale correlation function measured from the luminous red
galaxies sample of the Sloan Digital Sky Survey (SDSS) at $z_{\rm
b}=0.35$ \cite{Tegmark,Eisenstein}. Using the BAO data, one can
minimize the $\chi_{\rm BAO}^{2}$ defined as
\cite{Tegmark,Eisenstein},
\begin{equation}
\chi_{\rm BAO}^{2}=\frac{\left[A_{\rm obs}-A_{\rm
th}\right]^{2}}{\sigma_{A}^{2}},
\end{equation}
where
\begin{equation}
A_{\rm th}=\sqrt{\Omega_{m_0}}~E(z_{\rm
b};\textbf{p})^{-1/3}\left[\frac{1}{z_{\rm b}}\int_{0}^{z_{\rm
b}}\frac{dz'}{E(z';\textbf{p})}\right]^{2/3},\label{A-BAO}
\end{equation}
is the theoretical distance parameter. Here $A_{\rm
obs}=0.469(n_{s}/0.98)^{-0.35}\pm 0.017$ is measured from the SDSS
data \cite{Eisenstein} and the scalar spectral index $n_{s}$ is
taken to be 0.968, which has been updated from the WMAP7 data
\cite{Komatsu}.

\subsection{CMB shift parameter}

The structure of the anisotropies of the CMB radiation depends on
two eras in cosmology, i.e., the last scattering era and today. They
can also be applied to limit the cosmological models by minimizing
\begin{equation}
\chi_{\rm CMB}^{2}=\frac{\left[R_{\rm obs}-R_{\rm
th}\right]^{2}}{\sigma_{R}^{2}}.
\end{equation}
Here the shift parameter $R$ of the CMB is related to the position
of the first acoustic peak in the power spectrum of the temperature
anisotropies and given by \cite{YWang,Bond}
\begin{equation}
R_{\rm th}=\sqrt{\Omega_{m_0}}\int_{0}^{z_{\rm
rec}}\frac{dz'}{E(z';\textbf{p})},\label{R-CMB}
\end{equation}
where $z_{\rm rec}\simeq1091.3$ is the redshift at the recombination
epoch \cite{Komatsu}. Also the observational value of $R_{\rm obs}$
has been updated to $1.725\pm0.018$ from the WMAP7 data
\cite{Komatsu}.

\subsection{Observational Hubble Data (OHD)}

We use the compilation of Hubble parameter measurements estimated
with the differential evolution of passively evolving early-type
galaxies as cosmic chronometers. For the Hubble parameter
\begin{equation}
H(z)=-\frac{1}{1+z}\frac{{\rm d}z}{{\rm d}t},
\end{equation}
if ${\rm d}z/{\rm d}t$ is known, $H(z)$ is obtained directly
\cite{Hz}. Observed values of $H(z)$ can be used to estimate the
free parameters of the model by minimizing the quantity \cite{OHD}
\begin{equation}
\chi_{\rm OHD}^{2}=\sum_{\rm i=1}^{15}\frac{\left[H_{\rm obs}(z_{\rm
i})-H_{\rm th}(z_{\rm i},\textbf{p})\right]^{2}}{\sigma_{\rm
i}^{2}},
\end{equation}
where $\sigma_{\rm i}^{2}$ are the measurement variances. The 12
observational data of Hubble parameter \cite{Stern}-\cite{Riess} are
listed in Table \ref{H12}. We also use three more additional data:
$H(z = 0.24) = 79.69 \pm 2.32$, $H(z = 0.34) = 83.8 \pm 2.96$, and
$H(z = 0.43) = 86.45 \pm 3.27$ given by \cite{Gaztannaga}.

As the relative likelihood function is defined by ${\mathcal
L}=e^{-(\chi^2_{\rm total}-\chi^2_{\rm min})/2}$ \cite{nesseris},
the best-fit value of the model parameters follows from minimizing
the sum
\begin{equation}
\chi_{\rm total}^{2}=\tilde{\chi}_{\rm SN}^{2}+\chi_{\rm
BAO}^{2}+\chi_{\rm CMB}^{2}+\chi_{\rm OHD}^{2}.
\end{equation}

\section{Numerical results}\label{numeric}

Now, we discuss the constraints on model parameters of PLECH
$f(T)$-gravity model (\ref{fHDE}) by using the recent observational
data including SNeIa, CMB, BAO and OHD.

\begin{figure}
\includegraphics{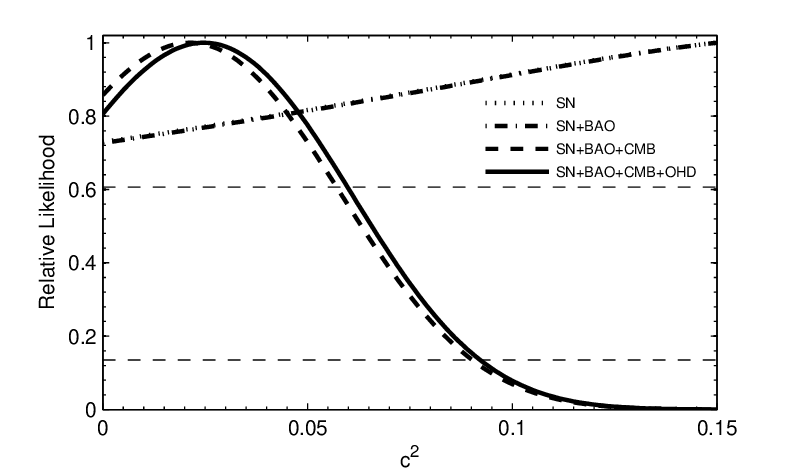}
      \vspace{5.5cm}
\caption[]{Same as Fig. \ref{alpha} for the parameter $c^2$.}
         \label{gamma}
   \end{figure}

\begin{figure}
\includegraphics{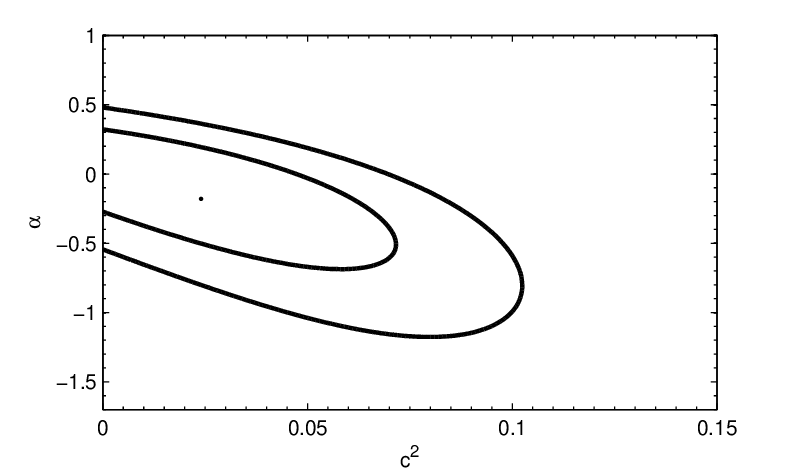}
      \vspace{5.5cm}
\caption[]{The $68.3\%~(1\sigma)$ and $95.4\% ~(2\sigma)$ CL
contours for $\alpha$ versus $c^2$ from the full data sets.}
         \label{contour}
   \end{figure}

The results are summarized in Table \ref{best-fit}, where we also
list the best-fit value of the corresponding parameter of the
$\Lambda$CDM model for comparison. At 68.3\% and 95.4\% confidence
levels (CLs), we obtain the best-fit value
$\alpha=-0.18^{+0.27}_{-0.34}(1\sigma) ^{+0.48}_{-0.77}(2\sigma)$
and $c^2 =
0.025^{+0.035}_{-0.025}(1\sigma)^{+0.067}_{-0.025}(2\sigma)$ for the
full data sets including SNeIa+CMB+BAO+OHD. The total $\chi^2$ of
the best-fit value of the PLECH $f(T)$-gravity model is $\chi_{\rm
min}^{2}=571.026$ for the full data sets with degrees of freedom
(dof) $=597$. The reduced $\chi^2$ is $0.956$, which is acceptable,
but $\chi_{\rm min}^{2}$ is smaller than the one for the
$\Lambda$CDM model, $\chi_{\Lambda\rm CDM}^{2}=571.306$ and
$\Omega_{m_0}=0.272^{+0.014}_{-0.012}(1\sigma)^{+0.027}_{-0.025}(2\sigma)$,
for the same data sets. The marginalized relative likelihood
functions ${\mathcal L}(\alpha)$ and ${\mathcal L}(c^2)$ are shown
in Figs. \ref{alpha} and \ref{gamma}, respectively. Figure
\ref{contour} shows the constraint on the PLECHDE parameter space
$\alpha - c^2$ at $1\sigma$ and $2\sigma$ CLs, using the full data
sets.

\begin{figure}
\includegraphics{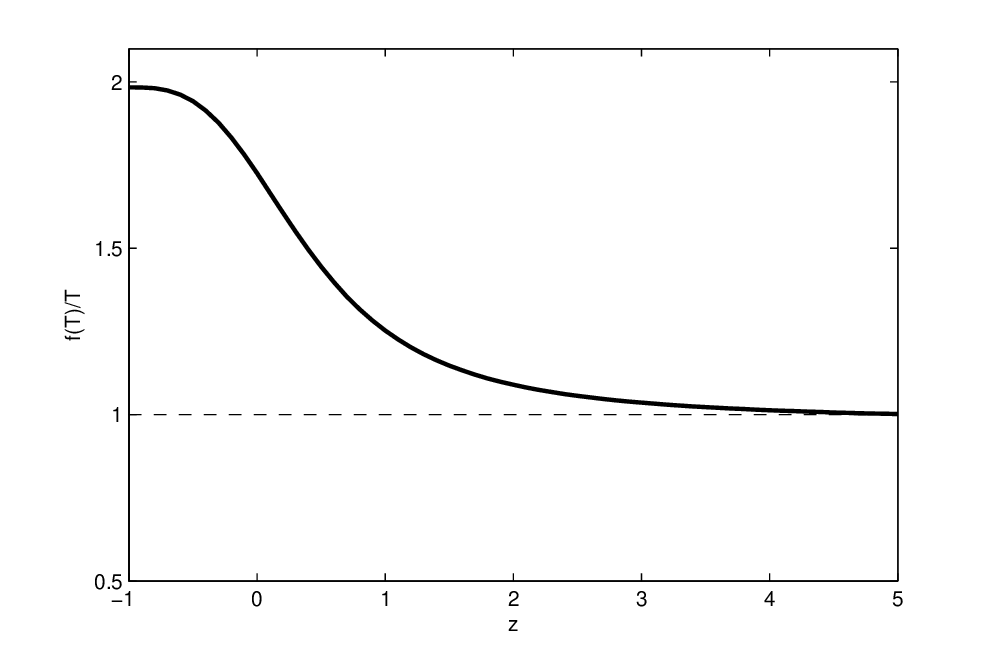}
      \vspace{5.2cm}
\caption[]{The evolution of PLECH $f(T)$-gravity model, Eq.
(\ref{fHDE}), versus $z$. The dashed line denotes the model $f(T)=T$
corresponding to the case of TG for comparison.}
         \label{f_T}
   \end{figure}

The evolution of the PLECH $f(T)$-gravity model, Eq. (\ref{fHDE}),
versus $z$ is shown in Fig. \ref{f_T}, where we also plot $ f(T)=T$
corresponding to the case of TG for comparison. Figure \ref{f_T}
shows that the PLECH $f(T)$-gravity model (\ref{fHDE}) satisfies the
condition
$$\lim_{|T|\rightarrow\infty}f/T\rightarrow 1,$$ at high redshift
which is compatible with the primordial nucleosynthesis and CMB
constraints \cite{WufT}.

\begin{figure}
\includegraphics{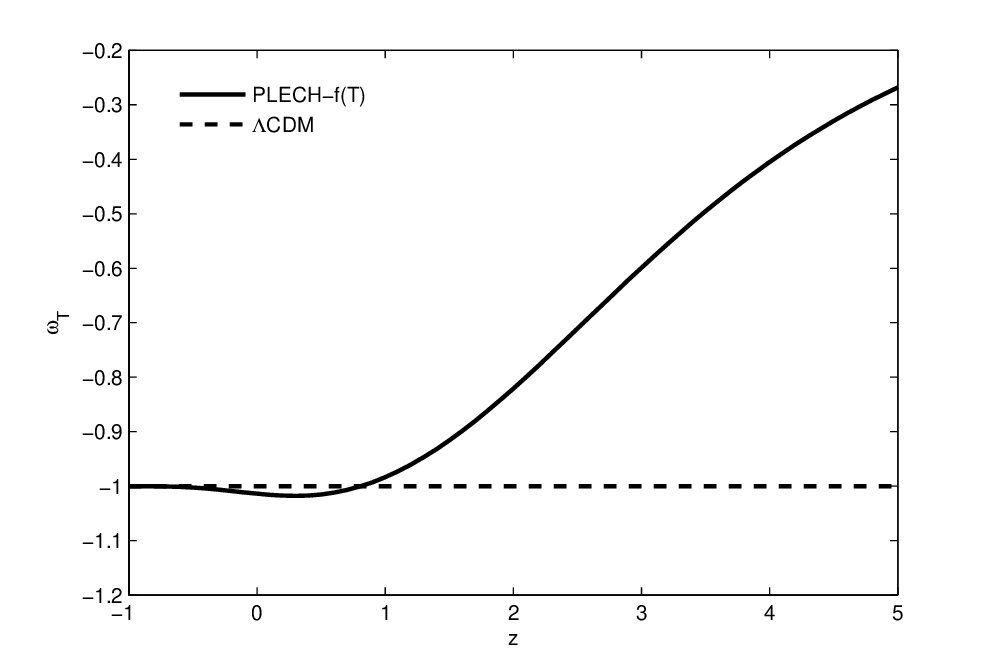}
      \vspace{5.9cm}
\caption[]{The effective torsion EoS parameter of the PLECH
$f(T)$-gravity model, Eq. (\ref{wHDE}) and $\Lambda$CDM using the
full data sets.}
         \label{w_eff}
   \end{figure}

The time evolution of the EoS parameter (\ref{wHDE}) for the
best-fit values of model parameters is plotted in Fig. \ref{w_eff}.
It shows that at early time ($z>>1$) we have
$\omega_T\rightarrow-0.3$ and at late time ($z=-1$) we get
$\omega_T\rightarrow-1$ which acts like the $\Lambda$CDM model. Also
at present time we have $\omega_{T_0}=-1.01$ which is in good
agreement with the recent observational result
$\omega_{T_0}=-0.93\pm 0.13~(68\%~\rm CL)$ deduced from the WMAP7
data \cite{Komatsu}. Figure \ref{w_eff} clears that the EoS
parameter of the PLECH $f(T)$-gravity model crosses the phantom
divide line from the values greater than $-1$ (quintessence phase)
to smaller than $-1$ (phantom phase) at $z=0.81$ which is compatible
with observations \cite{Sahni}.

\begin{figure}
\includegraphics{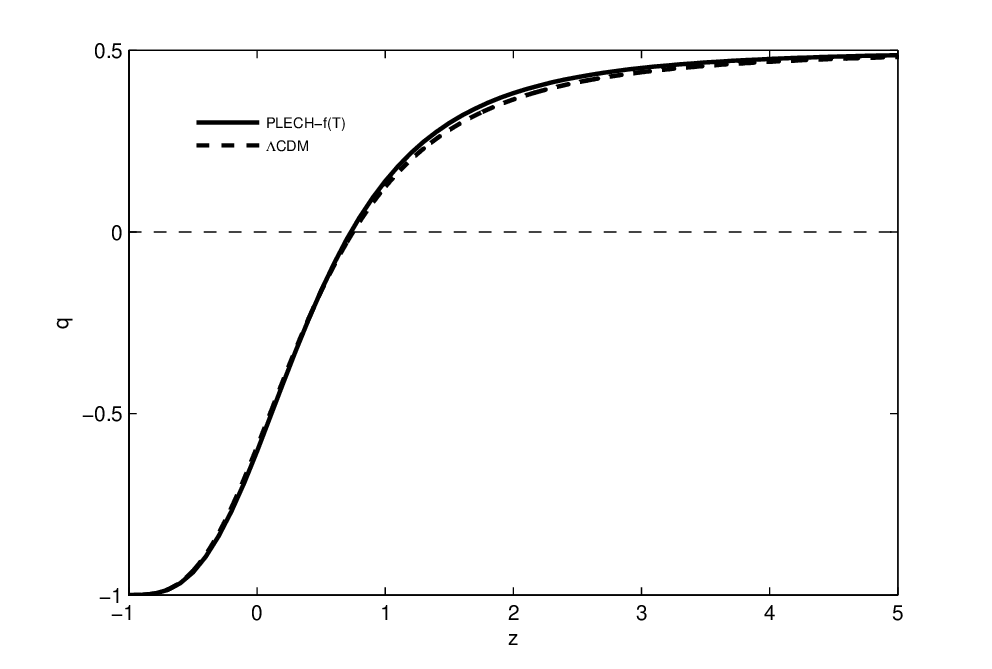}
      \vspace{5.9cm}
\caption[]{The best-fit of the deceleration parameter $q(z)$ of the
universe for the PLECH $f(T)$-gravity model, Eq. (\ref{q3}), and the
$\Lambda$CDM model using the full data sets.}
         \label{q}
   \end{figure}

In Fig. \ref{q}, we plot the evolutionary behavior of the
deceleration parameter of the universe, Eq. (\ref{q3}), with the
best-fit values of the PLECH $f(T)$-gravity and $\Lambda$CDM models.
Figure \ref{q} shows that very similar to the $\Lambda$CDM model the
universe transits from an early matter dominant regime to the de
Sitter phase in the future, as expected. The accelerating expansion
begins at transition redshift $z_{\rm t}=0.73$, which is later than
what the $\Lambda$CDM model predicts, $z_{\rm t}^{\Lambda\rm
CDM}=0.75$. The current best fit value of the deceleration parameter
in the PLECH $f(T)$-gravity model is obtained as $q_0=-0.6$ which
indicates the expansion rhythm of the current universe. This is in
agreement with the recent observational constraint
$q_0=-0.43_{-0.17}^{+0.13}~(68\%~\rm CL)$ obtained by the
cosmography \cite{CapCosmo}.

\begin{figure}
\includegraphics{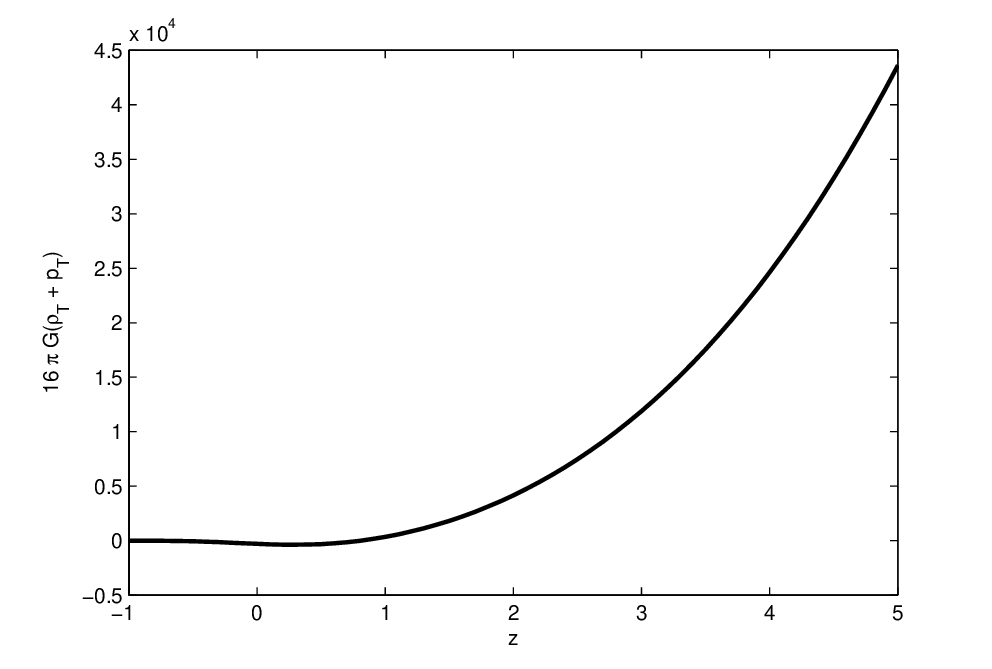}
      \vspace{5.9cm}
\caption[]{The evolution of $\rho_T + p_T$ versus $z$.}
         \label{NEC}
   \end{figure}

\begin{figure}
\includegraphics{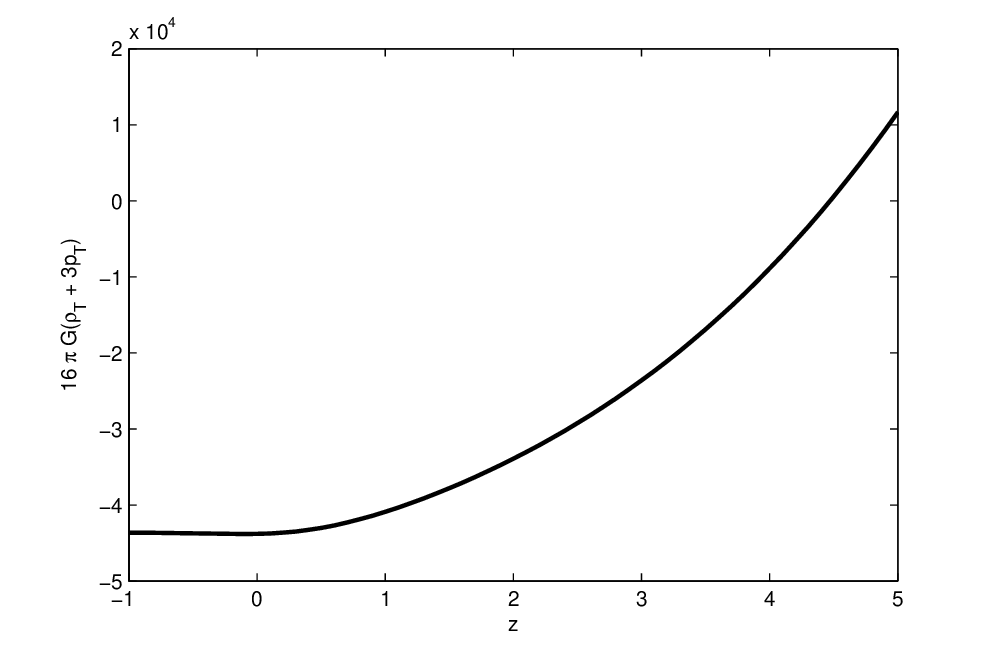}
      \vspace{5.9cm}
\caption[]{Same as Fig. \ref{NEC} for $\rho_T + 3p_T$.}
         \label{SEC}
   \end{figure}

\begin{figure}
\includegraphics{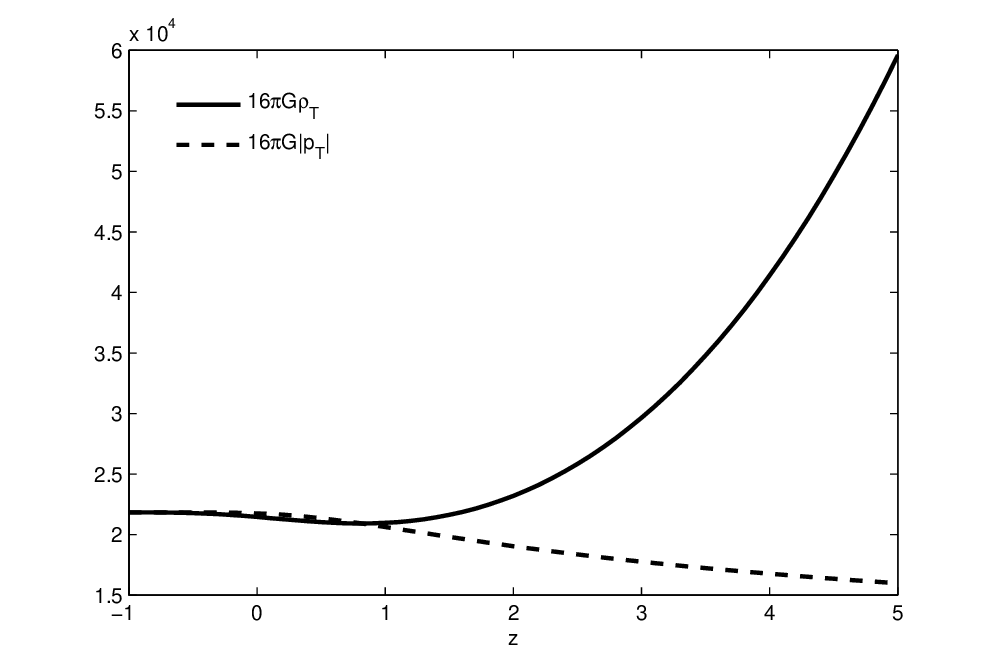}
      \vspace{5.9cm}
\caption[]{Same as Fig. \ref{NEC} for $\rho_T$ and $|p_T|$.}
         \label{DEC}
   \end{figure}

The evolutions of $\rho_{T} + p_{T}$, $\rho_{T}+3p_{T}$ and
($\rho_{T}$, $|p_{T}|$) versus $z$ are plotted in Figs. \ref{NEC},
\ref{SEC} and \ref{DEC}, respectively. Figures show that: (i) the
null energy condition (NEC), i.e. $\rho_T + p_T\geq 0$, is violated
when $z<0.81$ (see Fig. \ref{NEC}). (ii) The strong energy condition
(SEC), i.e. $\rho_{T} + p_{T}\geq 0$ and $\rho_{T}+3p_{T}\geq 0$, is
violated when $z<4.47$ (see Fig. \ref{SEC}). (iii) The weak energy
condition (WEC), i.e. $\rho_{T} + p_{T}\geq 0$ and $\rho_{T}\geq 0$,
is violated when $z<0.81$ (see Fig. \ref{DEC}). (iv) The dominant
energy condition (DEC), i.e. $\rho_{T}\geq 0$ and $\rho_{T}\geq
|p_{T}|$, is violated when $z<0.81$ (see Fig. \ref{DEC}).

\section{Cosmographic analysis}\label{cosmo}

Here we use the cosmographic constraints to check the viability of
$f(T)$ model without the need of explicitly solving the field
equations and fitting the data \cite{CapCosmo}. From Eqs.
(\ref{delta}) and (\ref{ep2}), $\gamma$ and $\epsilon$ are known.
This yields the $f_i=f^{(i)}(T_0)/(6H_0^2)^{-(i-1)}$ values, given
by Eqs. (4.23)-(4.26) in \cite{CapCosmo}, for $i=(2,3,4,5)$ where
$f^{(i)}(T)={\rm d}^if/{\rm d}T^i$ to be expressed as function of
$\alpha$ and $c^2$ when we fix $\Omega_{m_0}=0.1352/h^2$ from the
WMAP7 data. Following \cite{CapCosmo} for each $f_2$ and $f_3$
values of the sample obtained above from the cosmographic parameters
analysis, we solve $\hat{f}_2(\alpha,c^2)=f_2$ and
$\hat{f}_3(\alpha,c^2)=f_3$ to derive $\alpha$ and $c^2$ and
estimate the theoretically expected values for the other derivatives
$(f_4,f_5)$. The median is obtained as
\begin{eqnarray}
&&f_4=1.581\nonumber\\
&&f_5=5.456.\label{eqf41}
\end{eqnarray}
The above values for $(f_4,f_5)$ take place in the 68\% CL in Table
II in \cite{CapCosmo}. Hence, we conclude that the PLECH
$f(T)$-gravity model passes the cosmographic test.

\section{Generalized second law of thermodynamics (GSL)}\label{GSLaw}

\begin{figure}
\includegraphics{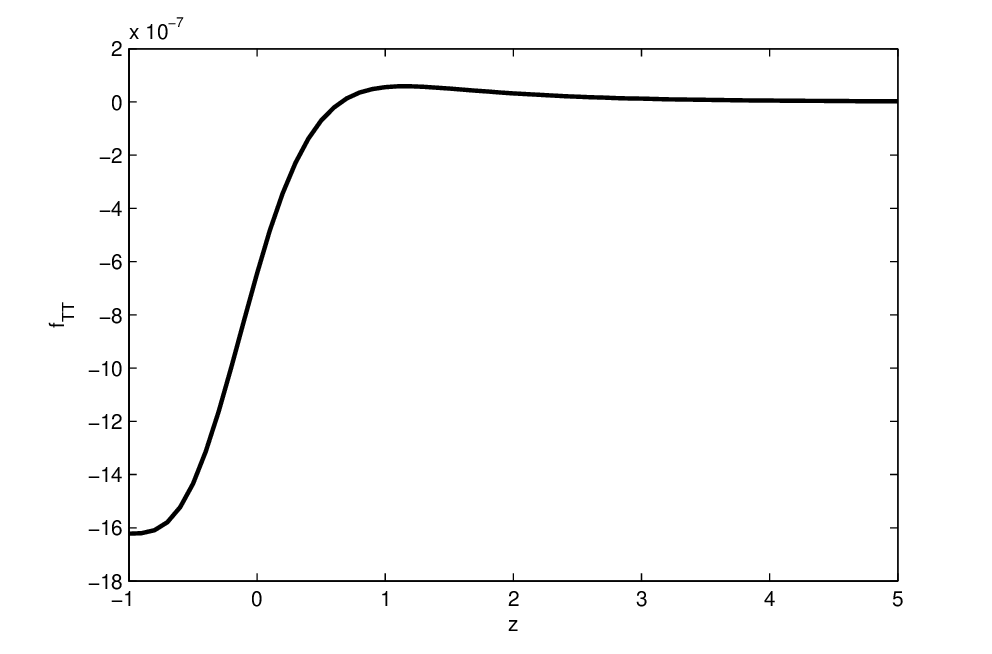}
      \vspace{5.5cm}
\caption[]{$f_{TT}$ versus $z$ for the PLECH $f(T)$-gravity model
(\ref{fHDE}).}
         \label{fTT}
   \end{figure}

Here, we investigate the validity of the GSL of gravitational
thermodynamics for PLECH $f(T)$-gravity model. Within the framework
of $f(T)$-gravity, the GSL is given by \cite{KA}
\begin{equation}
T_{\rm A}\dot{S}_{\rm
tot}=\frac{9}{8G}\left(\frac{f-2Tf_T}{f_T+2Tf_{TT}}\right)\left[
4f_{TT}+\left(\frac{f-2Tf_T}{f_T+2Tf_{TT}}\right)\left(\frac{f_T+5Tf_{TT}}{T^2}\right)\right]
,\label{TASdot}
\end{equation}
where $S_{\rm tot}=S_m+S_{\rm A}$ is the total entropy due to
contributions of both the matter and horizon. Also $T_{\rm A}$ is
the Hawking temperature \cite{Cai05}. In $f(T)$-gravity, the horizon
entropy
\begin{equation}
S_A=\frac{Af_T}{4G},\label{SA}
\end{equation}
where $A=4\pi \tilde{r}_{\rm A}^2$, is valid only when $f_{TT}$ is
small \cite{Miao}. We plot $f_{TT}$ versus $z$ for our model
(\ref{fHDE}) in Fig. \ref{fTT} which shows that the $f_{TT}$ is very
small. Hence Eq. (\ref{SA}) is valid for PLECH $f(T)$-gravity model.

\begin{figure}
\includegraphics{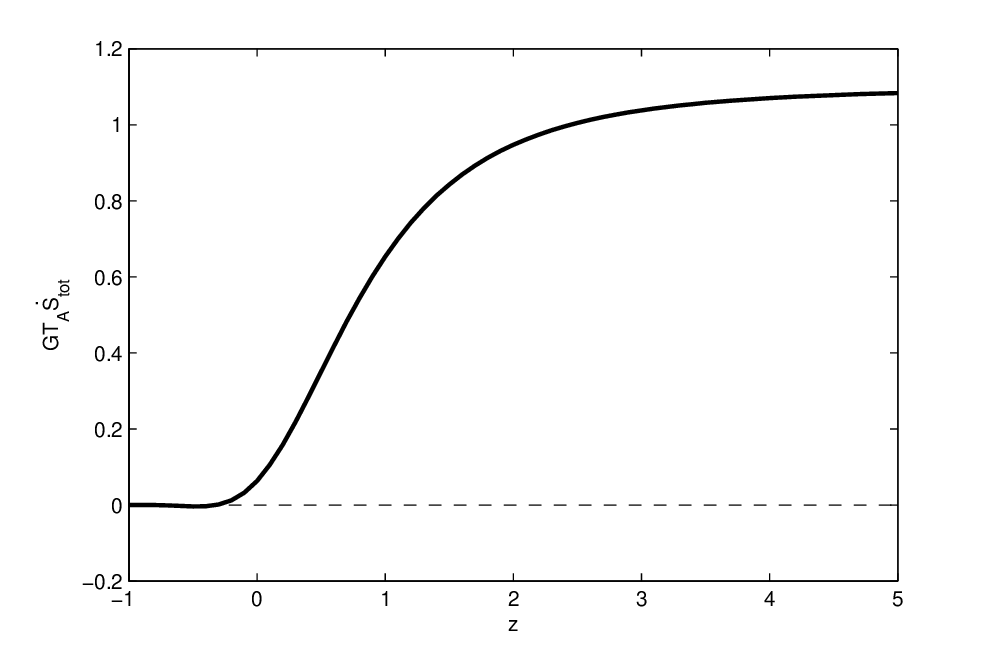}
      \vspace{5.5cm}
\caption[]{The variation of the GSL, Eq. (\ref{TSdotHDE}), versus
$z$ for model (\ref{fHDE}).}
         \label{GSL}
   \end{figure}

The GSL, Eq. (\ref{TASdot}), for the PLECH $f(T)$-gravity model
(\ref{fHDE}) reads
\begin{eqnarray}
GT_{A}\dot{S}_{\rm tot}=\frac{\rm
I}{8(\alpha-1)\left[2(c^2-1)T+\alpha\gamma(-T)^{\alpha/2}\right]^2T},\label{TSdotHDE}
\end{eqnarray}
where
\begin{eqnarray}
{\rm I}=&-&9\left[(c^2-1)T+\gamma(-T)^{\alpha/2}\right]\left\{
 4(\alpha-1)(c^2-1)^2T^2\right.~~~~~~~~~~~~~~~~~~~~~~~~~~~~~~~~~~~~~~~~~~~~~~~~~~~
 \nonumber\\&+&\left.(c^2-1)\left[(-4+\alpha(4+\alpha))\gamma(-T)^{\alpha/2}-(\alpha-1)\epsilon\sqrt{-T}\right]T\right.
 \nonumber\\&+&\left.\gamma\left[-\alpha\gamma(8+\alpha(-9+2\alpha))(-T)^{\alpha/2}+\epsilon(-3+5\alpha-2\alpha^2)\sqrt{-T}\right](-T)^{\alpha/2}
  \right\}.
\end{eqnarray}
The variation of the GSL (\ref{TSdotHDE}) versus $z$ is plotted in
Fig. \ref{GSL}. Figure shows that the GSL for our model is satisfied
from the early times to the present epoch. But in the future the GSL
is violated for $z<-0.32$. These are in good agreement with those
obtained by \cite{KS} for the power-law corrected entropy-area
relation (\ref{ec}).

\section{The growth of structure formation}\label{str}

Here, we investigate the growth rate of matter density perturbation
in PLECH $f(T)$-gravity model. In $f(T)$-gravity, the matter density
contrast $\delta_m=\delta\rho_m/\rho_m$ satisfies \cite{Rui Zheng}
\begin{equation}
\ddot{\delta}_{m}+2H\dot{\delta}_{m}-4{\pi}G_{\rm
eff}\rho_{m}\delta_{m}=0,\label{eqdelta}
\end{equation}
where $G_{\rm eff}=\frac{G}{f_T}$ is the effective Newton's
constant. Now we define a new variable $g(a)$, namely
$g(a)\equiv\delta_{m}/a$ which does not depend on $a$ during the
matter era. Thus the initial conditions are $g(a_{\rm i})=1$ and
$\frac{{\rm d}g}{{\rm d}\ln a}\mid_{a=a_{\rm i}}=0$, where $a_{\rm
i}=1/31$ (i.e., $z=30$) \cite{Rui Zheng}. Equation (\ref{eqdelta})
in terms of $g(a)$  becomes
\begin{equation}
\frac{{\rm d}^2g}{{\rm
d}\ln{a^2}}+\left(4+\frac{\dot{H}}{{H}^{2}}\right)\frac{{\rm
d}g}{{\rm d}\ln{a}}
+\left(3+\frac{\dot{H}}{{H}^{2}}-\frac{4{\pi}G_{\rm eff}\rho_{\rm
m}}{H^2}\right)g=0.\label{eqg1}
\end{equation}
From Eqs. (\ref{q1}), (\ref{q2}), (\ref{fHDE}), (\ref{delta}) and
(\ref{ep2}) one can get
\begin{equation}
\frac{\dot{H}}{H^2}=-\frac{3}{2}\left(\frac{2f_T-f/T}{f_T+2Tf_{TT}}\right)=
-\frac{3}{2}\left[\frac{1-{c}^{2}+(\Omega_{m_{0}}+c^{2}-1)E^{\alpha-2}}
{1-{c}^{2}+\frac{\alpha}{2}(\Omega_{m_{0}}+c^{2}-1)E^{\alpha-2}}\right],\label{eqHdotH2}
\end{equation}
where $E(z)=H(z)/H_0$ is given by Eq. (\ref{eqE}). Also with the
help of Eq. (\ref{Hdot}) and (\ref{eqHdotH2}) one can obtain
\begin{equation} \frac{4{\pi}G_{\rm eff}\rho_{m}}{H^2}=
\frac{3}{2}\left(2-\frac{f}{Tf_T}\right)=\frac{3}{2}
\left[\frac{(1-{c}^{2})E+(\Omega_{m_{0}}+c^{2}-1){E}^{\alpha-1}}
{c^{2}+(1-c^{2})E+\frac{1}{2}\Big(\frac{\alpha}{\alpha-1}\Big)(\Omega_{m_{0}}+c^{2}-1)(E^{\alpha-1}-1)}\right].\label{eqG}
\end{equation}
Taking the derivative of Eq. (\ref{eqE}) with respect to $\ln a$
gives
\begin{equation}
\frac{{\rm d}\ln E}{{\rm
d}\ln{a}}=-\frac{3}{2}\left[\frac{1-c^2+(\Omega_{m_{0}}+c^2-1)E^{\alpha-2}}{1-c^2+\frac{\alpha}{2}
(\Omega_{m_{0}}+c^2-1)E^{\alpha-2}}\right],\label{eqh2}
\end{equation}
with the initial condition $E(a=1)=1$. Note that to obtain the
evolutionary behavior of $g(a)$, we need to solve Eqs. (\ref{eqg1})
and (\ref{eqh2}), numerically.

\begin{figure}
\includegraphics{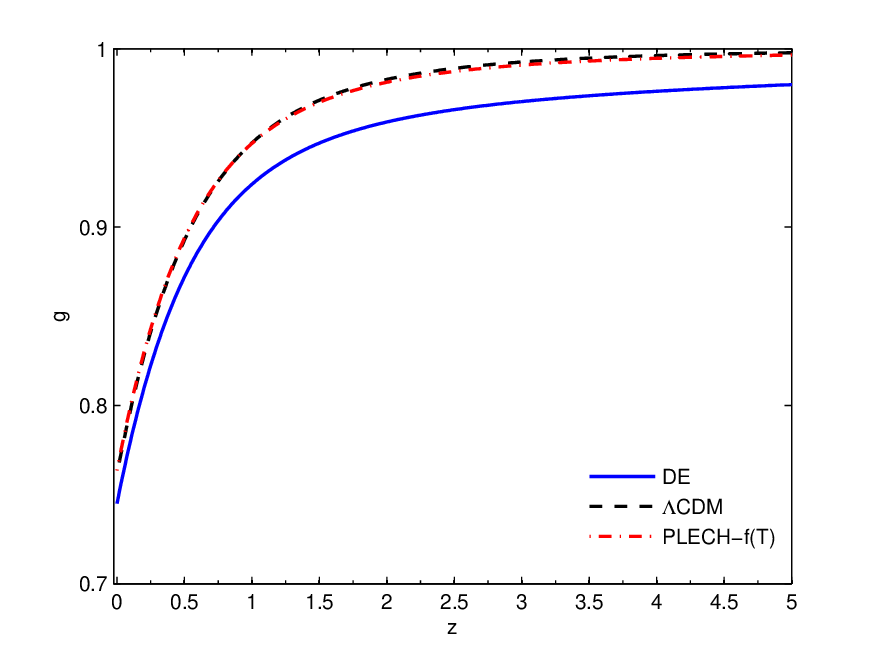}
      \vspace{6.3cm}
\caption[]{The variation of $g(a)\equiv\delta_{m}/a$ versus $z$ in
the PLECH $f(T)$-gravity model, ${\Lambda}$CDM model, and DE model
with the same EoS in GR.}
         \label{g}
   \end{figure}

In Fig. \ref{g}, we plot the evolutionary behavior of the
$g(a)=\delta_{m}/a$ with the best-fitting values of the PLECH
$f(T)$-gravity model, ${\Lambda}$CDM model, and DE model with the
same EoS in GR. Note that the variation of $g(a)$ for the DE
scenario in GR with the same EoS as the effective DE in $f(T)$ is
obtained by replacing $G_{\rm eff}$ with $G$ in Eq. (\ref{eqg1}).
Figure \ref{g} shows that: (i) $g(a)$ for the three models starts
from an early matter dominant phase, i.e. $g\simeq 1$, and decreases
during history of the universe. (ii) For a given $z$, $g(a)$ in the
PLECH $f(T)$-gravity model like the ${\Lambda}$CDM model gets
greater than that in the DE model with the same EoS in GR.

\begin{figure}
\centering\includegraphics{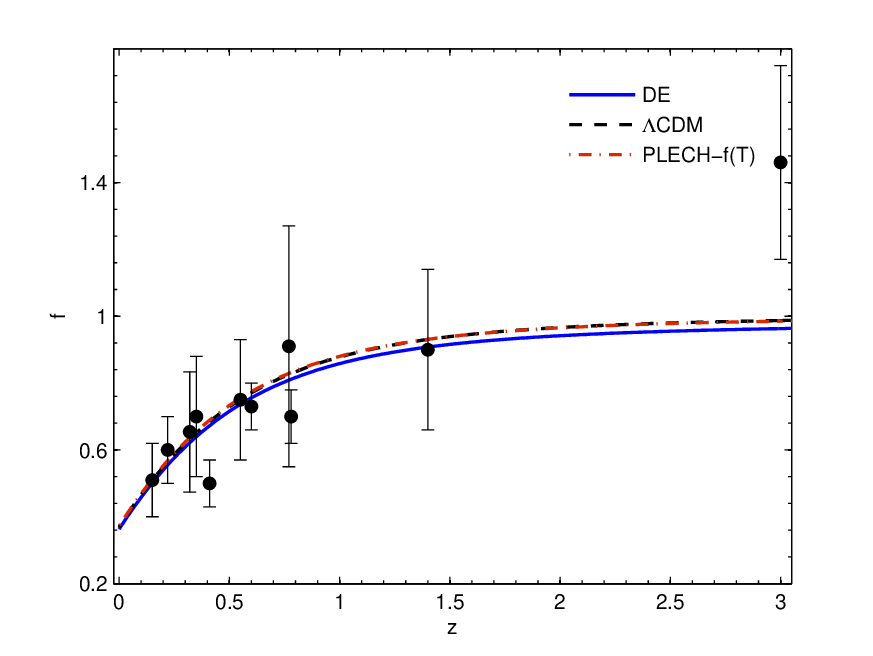}
      \vspace{6.3cm}
\caption[]{Evolution behaviors of the growth factor for the PLECH
$f(T)$-gravity model, ${\Lambda}$CDM model, and DE model with the
same EoS in GR.}
         \label{f}
  \end{figure}

Figure \ref{f} shows the evolutionary behavior of the growth factor
$f(z)$ defined as \cite{Peebles}
\begin{equation}
f(z)=\frac{{\rm d}\ln\delta_{m}}{{\rm d}\ln a}=-(1+z)\,\frac{{\rm
d}\ln\delta_{m}}{{\rm d}z},
\end{equation}
with the best-fitting values of the PLECH $f(T)$-gravity model,
${\Lambda}$CDM model, and DE model with the same EoS in GR. The
growth factor data are listed in Table \ref{fdata}. Figure \ref{f}
shows that the PLECH $f(T)$-gravity model fits the data of the
growth factor well as the $\Lambda$CDM model.

\section{Conclusions}\label{conc}

Within the framework of $f(T)$ modified teleparallel theory, we
reconstructed a $f(T)$ model according to the PLECHDE model. We
fitted the model with current observational data, including SNeIa,
BAO, CMB and OHD. We obtained the constraint results of PLECH
$f(T)$-gravity model parameters,
$\alpha=-0.18^{+0.27}_{-0.34}(1\sigma) ^{+0.48}_{-0.77}(2\sigma)$
and $c^2 =
0.025^{+0.035}_{-0.025}(1\sigma)^{+0.067}_{-0.025}(2\sigma)$ for the
full data sets. The minimal $\chi^2$ gives $\chi_{\rm
min}^{2}=571.026$ with dof$=597$. The reduced $\chi^2$ equals to
$0.956$ which is acceptable. The $\chi_{\rm min}^{2}$ is slightly
smaller than the one for the $\Lambda$CDM model, $\chi_{\Lambda\rm
CDM}^{2}=571.306$, for the same data sets. Using the best-fit values
of the PLECH $f(T)$-gravity model parameters, we also studied the
evolutionary behaviors of the effective torsion EoS parameter of
PLECH $f(T)$-gravity model, the deceleration parameter of the
universe and different energy conditions.

Using a cosmographic analysis approach, we also checked the
viability of our model without the need of explicitly solving the
field equations and fitting the data. We further examined the
validity of the GSL of gravitational thermodynamics for the PLECH
$f(T)$-gravity model. Finally, we pointed out the growth of
structure formation in our model. Our results show the following.

(i) The condition $f/T\rightarrow 1$ is satisfied for our model at
high redshift ($|T|\rightarrow\infty$) which is compatible with the
primordial nucleosynthesis and CMB constraints.

(ii) The effective torsion EoS parameter $\omega_T$ varies from
$\omega_T>-1$ to $\omega_T=-1$. At late time, it behaves like the
$\Lambda$CDM model. For the present time, we obtain
$\omega_{T_0}=-1.01$ which acts like phantom universe and it is in
good agreement with the recent observational result deduced from the
WMAP7 data \cite{Komatsu}. Also $\omega_T$ shows a transition from
the quintessence phase ($\omega_T>-1$) to the phantom regime
($\omega_T<-1$) at $z=0.81$ which is compatible with observations
\cite{Sahni}.

(iii) The variation of the deceleration parameter $q$ shows that the
universe transits from an early matter dominant epoch ($q=0.5$) to
the de Sitter era ($q=-1$) in the future, as expected. The
accelerating expansion begins at transition redshift $z_{\rm
t}=0.73$, which is later than what the $\Lambda$CDM model predicts,
$z_{\rm t}^{\Lambda\rm CDM}=0.75$. The deceleration parameter
$q_0=-0.6$ obtained at the present is compatible with the recent
observational constraint obtained by the cosmography
\cite{CapCosmo}.

(iv) The NEC, WEC and DEC are violated for $z<0.81$ when the
universe enters the phantom phase. Also the SEC does not hold for
$z<4.47$.

(v) Cosmographic analysis shows that the PLECH $f(T)$-gravity model
is favored by the observational data.

(vi) The GSL of gravitational thermodynamics holds for our $f(T)$
model from the early times to the present epoch. But in the future,
the GSL is violated for $z<-0.32$.

(vii) The evolution of the growth factor in the PLECH $f(T)$-gravity
model shows that our model like the $\Lambda$CDM model fit the data
very well.

\subsection*{Acknowledgements}

The authors thank the referee for his/her valuable comments. The
works of K. Karami and Z. Safari have been supported financially by
Research Institute for Astronomy and Astrophysics of Maragha (RIAAM)
under research project No. 1/2782-44.

\clearpage
 \begin{figure}
\includegraphics{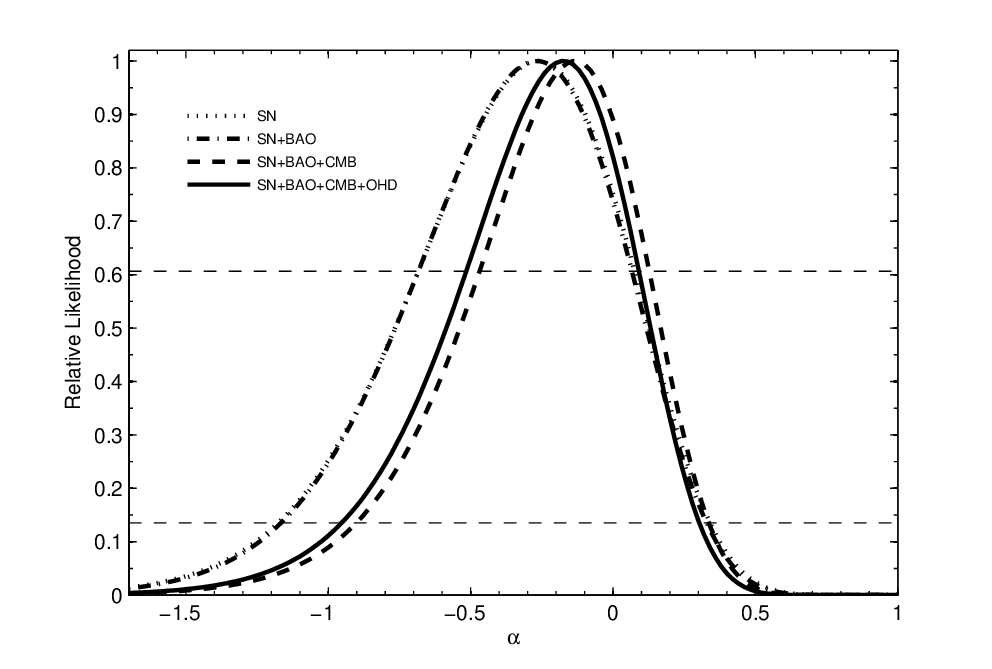}
      \vspace{5.5cm}
\caption[]{The 1D marginalized likelihood of $\alpha$. The
horizontal dashed lines give the bounds with $1\sigma$ and $2\sigma$
CLs.}
         \label{alpha}
   \end{figure}
\clearpage
\clearpage
\clearpage
\clearpage
\begin{table}\small
\centering \caption{The observational $H(z)$ data
\cite{Stern}-\cite{Riess}.}
\begin{tabular}{lcccccccccccc}\hline\noalign{\smallskip}
$z$ & $0$ & $0.1$  & $0.17$ & $0.27$ & $0.4$ & $0.48$ & $0.88$ &
$0.9$ & $1.30$ & $1.43$ & $1.53$ & $1.75$
\\\hline\noalign{\smallskip}
 $H(z)$ &  $74.2$ &$69$ & $83$ & $77$ & $95$ & $97$ &
$90$  &$117$& $168$ & $177$ & $140$ & $202$
\\$1\sigma$ & $\pm3.6$ &$\pm12$ & $\pm 8$ & $\pm14$ &
$\pm17$ & $\pm60$ & $\pm40$ & $\pm23$ & $\pm17$ & $\pm18$ & $\pm14$
&$\pm 40$
\\
\hline
\end{tabular}
\label{H12}\\
\end{table}
\begin{table}
\centering\caption[]{The best-fit values of the parameters $\alpha$
and $c^2$ within the 68.3\% ($1\sigma$) and 95.4\% ($2\sigma$) CLs
for each observational data set for the PLECH $f(T)$-gravity model.
Columns 5, 6, and 7 show the current effective torsion EoS
parameter, the current deceleration parameter of the universe and
the transition redshift, respectively. The last row shows the
best-fit result of the $\Lambda$CDM model using the full data sets
for comparison.}
\begin{tabular}{lccccccc}\hline\noalign{\smallskip}
{\rm Data} & $\alpha$ &$c^2$ &$\chi^{2}_{\rm
min}$&$\omega_{T_0}$&$q_{0}$&$z_{\rm t}$&$\chi^{2}/dof$
\\\hline\noalign{\smallskip}
 $\rm SN$ & $-0.26_{-0.42 -0.91}^{+0.34 +0.60}$ & $0.15$ &
$562.259$&$-0.97$&$-0.5$&$0.63$&$0.941$\\\\
$\rm SN+BAO$ & $-0.26_{-0.42 -0.91}^{+0.33 +0.59}$ & $0.15$ & $562.362$ &$-0.97$& $-0.5$ & $0.63$&$0.942$ \\\\
$$\rm SN+BAO\\+CMB$$ & $-0.14_{-0.33 -0.76}^{+0.27 +0.48}$ &
$0.021^{+0.036 +0.069}$ &
$562.381$&$ -1.01 $&$-0.6$&$0.74$&$0.942$\\\\
$$SN+BAO+\\CMB+OHD$$ & $-0.18^{+0.27 +0.48}_{-0.34 -0.77}$ &
$0.025^{+0.035 +0.067}_{-0.025 -0.025}$ &
$571.026$&$-1.01$&$-0.6$&$0.73$&$0.956$\\\\
$\Lambda\rm CDM$ & $-$ & $-$ & $571.306$ & $-1$&$-0.6$ & $0.75$&$0.957$\\
\hline
\end{tabular}
\label{best-fit}
\end{table}
\begin{table}\small
\centering \caption{The observational data for the linear growth
rate $f_{\rm obs}(z)$.}
\begin{tabular}{lccccccccccc}\hline
$z$ & $0.15$ & $0.22$  & $0.32$ & $0.35$ & $0.41$ & $0.55$ & $0.60$
& $0.77$ & $0.78$ & $1.4$ & $3.0$
\\\hline
 $f_{\rm obs}$ &  $0.51$ &$0.60$ & $0.654$ & $0.70$ & $0.50$ & $0.75$ &
$0.73$  &$0.91$& $0.70$ & $0.90$ & $1.46$
\\\hline$1\sigma$ & $0.11$ &$0.10$ & $0.18$ & $0.18$ &
$0.07$ & $0.18$ & $0.07$ & $0.36$ & $0.08$ & $0.24$ & $0.29$
\\
\hline${\rm Ref.}$ & $\cite{Hawkins}$ &$\cite{Blake}$ &
$\cite{Reyes}$ & $\cite{Tegmarks}$ & $\cite{Blake}$ & $\cite{Ross}$
& $\cite{Blake}$ & $\cite{Guzzo}$ & $\cite{Blake}$ & $\cite{Angela}$
& $\cite{Donald}$\\\hline
\end{tabular}
\label{fdata}\\
\end{table}

\end{document}